\documentclass[12pt,epsf]{article}

\usepackage{amsmath}
\usepackage{amsthm, amssymb}
\usepackage{amsfonts}
\usepackage{fixltx2e}%
\usepackage{booktabs} % Better tables
\usepackage{url}
\usepackage{graphicx}

\usepackage{xcolor}
\usepackage{framed}
\definecolor{shadecolor}{gray}{0.95}

\usepackage{mathrsfs}
\usepackage{setspace}
\usepackage{subfigure}
\usepackage{slashed}
\usepackage{mathtools}
\usepackage{amssymb}
\usepackage{amsfonts}
\usepackage{tensind}
\usepackage{float}
\usepackage{microtype}%
\usepackage{amsmath}%
\usepackage{setspace}
\usepackage{amsthm}%
\mathtoolsset{showonlyrefs}
\usepackage[utf8]{inputenc}%
\usepackage{csquotes}
\usepackage{textcomp}%
\usepackage[english]{babel}%
\usepackage{empheq}

\usepackage{bm}

\usepackage{tikz}
\usetikzlibrary{decorations.markings}
\usetikzlibrary{positioning,arrows,matrix,patterns}
\usetikzlibrary{calc}
\usetikzlibrary{decorations.pathmorphing}
\usepackage{tikz-cd}

\newcommand{\D}{\,\mathrm{d}}

\DeclareMathAlphabet{\mathpzc}{OT1}{pzc}{m}{it}
\newcommand{\nbox}{{\,\lower0.9pt\vbox{\hrule \hbox{\vrule height 0.2 cm \hskip 0.19 cm \vrule height 0.2 cm}\hrule}\,}}

\def\href#1#2{#2}

\usepackage{color}

\usepackage{xcolor}

\usepackage[bookmarks,colorlinks,urlcolor=blue,colorlinks = true,
linkbordercolor = {white},
citebordercolor = {white},
citecolor = {blue},
linkcolor = {blue}]{hyperref} % Add hyperref  type links in the document, colors

\begin{document}
\begin{titlepage}
	
		\begin{flushright}
		
		\texttt{ MPP-2019-180 }
		
	\end{flushright}
\begin{NoHyper}
\hfill
\vbox{
    \halign{#\hfil         \cr
           } % end of \halign
      }  % end of \vbox
\vspace*{20mm}
\begin{center}
{\Large \bf Gravitational memory in the bulk}

\vspace*{15mm}
\vspace*{1mm}
Henk Bart 
\vspace*{1cm}
\let\thefootnote\relax

{
	\textit{Max-Planck-Institut f\"ur Physik,\\
		F\"ohringer Ring 6, 
		80805 M\"unchen, Germany\\}
	\quad \\
	\href{mailto:hgjbart@mpp.mpg.de}{hgjbart@mpp.mpg.de}
	
}

\vspace*{1cm}
%%\maketitle
\end{center}

\begin{abstract}
	A method for detecting gravitational memory is proposed. It makes use of ingoing null geodesics instead of timelike geodesics in the original formulation by Christodoulou. It is argued that the method is applicable in the bulk of a spacetime. In addition, it is shown that BMS symmetry generators in Newman-Unti gauge have an interpretation in terms of the memory effect. This generalises the connection between BMS supertranslations and gravitational memory, discovered by Strominger and Zhiboedov at null infinity, to the bulk.
\end{abstract}
\end{NoHyper}

\end{titlepage}
%\tableofcontents
\vskip 1cm
\begin{spacing}{1.15}

\section{Introduction}
In General Relativity, there exists the gravitational memory effect, which was discovered by Zel'dovich and Polnarev \cite{Zeldovich1974}, then studied by Braginsky and Thorne \cite{Braginsky1987,Thorne:1992sdb} in the linearised theory, and at null infinity by Christodoulou in the nonlinear theory in \cite{Christodoulou:1991cr}. It is a statement about how the relative distance between geodesics permanently changes after the passing of a burst of radiation. 

This effect is conceptually nontrivial in the following sense. Usually, one imagines that a ring of test particles subject to a gravitational plane wave oscillates in the $+$ or $\times$ polarisation directions, and then returns to its initial state. The gravitational memory effect states that this is not true; the relative distance between test particles of the ring is permanently changed after the passing of the wave. The effect is often referred to as the Christodoulou memory effect, because Christoudoulou made the observation\footnote{Blanchet and Damour \cite{Blanchet1992} independently obtained the same result.} that gravitational backreaction in the linearised theory cannot be ignored. 

Christodoulou formulated the memory effect at null infinity with the help of test particles on timelike geodesics that are initially at rest. This formulation relies on the fact that it is possible at null infinity to define a good notion of ``test particles initially at rest''. In this note, we shall be interested in formulating the memory effect elsewhere\footnote{See \cite{Shore:2018kmt} for a geometric approach that goes beyond the weak-field analysis. See \cite{Donnay:2018ckb,Hawking:2016sgy} for statements about the memory effect in black hole spacetimes.} in the spacetime. We have in mind a gravitational wave on a black hole background. One could try to set up a ring of timelike test particles around the black hole, wait for the wave to pass by, and measure the relative displacements. However, since timelike geodesics close to a black hole are gravitated inwards, it is difficult to interpret which part of the displacement can be explained in terms of the gravitational memory effect.

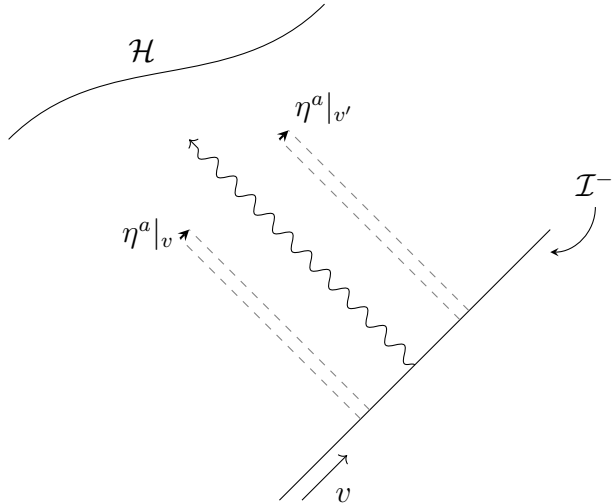
\begin{figure}
	\begin{center}
	\begin{tikzpicture}[ scale=0.6]
	\coordinate (i-) at (0,0) {};
	\coordinate (i0) at (6,6) {};
	\node (shift) at (1,1) {};
	\node (rd-shift) at (-1,1) {};

	\draw (i-)  --  (i0);
		\draw[->,>=stealth] ($(i0) + (1,0.5)$) to[out=-90,in=0] node[at start,above ]{$\mathcal{I}^-$} ($(i0) + (0,-0.5)$);
		\draw[->] ($(i-)+(0.5,0)$) -- node[midway, below right,rotate = 0] {$v$}($(i-)+(0.5,0)+(shift)$);
	
	\draw[dashed,color=gray] ($(i-)+2*(shift)$) -- ($(i-)+2*(shift)+4*(-1,1)$);
	\draw[dashed,color=gray] ($(i-)+1.8*(shift)$) --  ($(i-)+1.8*(shift)+4*(-1,1)$);
	
	\draw[dashed,color=gray] ($(i-)+4*(shift)$) -- ($(i-)+4*(shift)+4*(-1,1)$);
	\draw[dashed,color=gray] ($(i-)+4.2*(shift)$) -- ($(i-)+4.2*(shift)+4*(-1,1)$);
	
	\node[ left] (I)    at ($(i-)+1.9*(shift)+4*(-1,1)$)  {$\eta^a|_{v}$};
	\node[above right] (II)    at ($(i-)+4.1*(shift)+4*(-1,1)$)  {$\eta^a|_{v'}$};
	
	\draw[->, thick,>=stealth] ($(i-)+1.8*(shift)+4*(-1,1)$) -- ($(i-)+2*(shift)+4*(-1,1)$);
	\draw[->, thick,>=stealth] ($(i-)+4*(shift)+4*(-1,1)$) -- ($(i-)+4.2*(shift)+4*(-1,1)$);
	
%	\draw[->, thick,>=stealth] ($(i-)+1.8*(shift)-0.1*(rd-shift)$) -- ($(i-)+2*(shift)-0.1*(rd-shift)$);
%	\draw[->, thick,>=stealth] ($(i-)+4*(shift)-0.1*(rd-shift)$) -- ($(i-)+4.2*(shift)-0.1*(rd-shift)$);
	
%	\node[below right ] (Ia)    at ($(i-)+1.9*(shift)$)  {$d^a$};
%	\node[ below right] (IIa)    at ($(i-)+4.1*(shift)$)  {$d^a$};
	
	\draw[->,decorate, decoration={snake}] ($(i-)+3*(shift)$)  -- node[at start, right, rotate=-45,] {$ $} ($(i-)+3*(shift)+5*(-1,1)$);

	%%%----Black Hole---%%%
	\draw[ ] ($(i-)+8*(rd-shift)+(2,0)$)    to[out=45,in=-135] node [midway,  above left] {$\mathcal{H}$} ($(i-)+8*(rd-shift)+(2,0)+(7,3)$);
	
	\end{tikzpicture}

	\caption{A gravitational wave travels towards a (dynamical) black hole horizon $\mathcal{H}$. The passing of the wave changes the geodesic deviation $\eta^a$ of a (specific) pair of ingoing null geodesics. This geodesic deviation is compared at two different values of $v$, as a way of detecting gravitational memory in the bulk of the spacetime. }
	\end{center}
\label{fig:sketch}
\end{figure}

Therefore, we shall provide a different formulation of the  memory effect. Instead of timelike geodesics, we consider a pair of ingoing null geodesics. We measure the geodesic deviation between the light rays at their affine time\footnote{The affine parametrisation of the geodesics has to be chosen in a suitable manner, which will be explained later.} (or radius) $r$. At a later time, we introduce another pair of such light rays. Comparison of the geodesic deviation of the second pair to the first -- at the affine time $r$ -- is a method for detecting gravitational memory. The setup is sketched in \autoref{fig:sketch}.

One advantage of this method is clear: an external observer can wait as long as he or she wants to perform the measurement with the second pair of light rays, and still be able to measure the permanent displacement. This is not in general possible in the usual setup with timelike geodesics. 

Now, we come to the second point of this note. In \cite{Strominger:2014pwa} it was observed that the gravitational memory effect is related to the subject of  \emph{BMS symmetries}\footnote{These were defined in \cite{Newman:1962cia,Bondi:1962px}. See \cite{Strominger:2017zoo,Barnich:2011mi} for a review.} at null infinity.  The relation is that a change of the relative displacement between geodesics due to a burst of radiation can be understood as the action of a supertranslation. In the present note, we shall observe that our formulation of the gravitational memory effect in the bulk can be understood in terms of the action of BMS supertranslations in Newman-Unti gauge \cite{Newman:1962cia}. This generalises the connection \cite{Strominger:2014pwa} between BMS supertranslations and gravitational memory   to the bulk of the spacetime.

The organisation of this note is as follows. In \autoref{sec:preliminaries} we review null geodesic generators of null hypersurfaces. In \autoref{sec:memory} we provide a new formulation of the memory effect in terms of null geodesics. In \autoref{sec:BMS} we discuss the memory effect in relation to BMS symmetries.

\section{Null geodesics}\label{sec:preliminaries}
In this section we show how, starting from a null geodesic generator $n^a$ of a null hypersurface $\Sigma_v$, it is possible to construct neighbouring geodesics. In the remainder of this note we shall frequently consider the \emph{geodesic deviation} between the original null geodesic $n^a$ and its deformation.

Consider a spacetime $M$ with metric $g_{ab}$. Denote by $x^a(\tau)$ a path in $M$ parametrised by $\tau$. Then $x^a(\tau)$ is a geodesic when 
\begin{equation}
\frac{\partial H}{\partial p_a} = \dot{x}^a  \quad \text{and} \quad     \frac{\partial H}{\partial x^a} = -\dot{p}_a,
\end{equation}
where 
\begin{equation}
H = \frac12 g^{ab}(x) p_a p_b.
\end{equation}
That is,
\begin{align}
\dot{x}^a &=  g^{a b}p_b  \label{eq:geodesic-eqn1},\\
\dot{p}_a &= -\frac12 (\partial_a g^{bc} ) p_b p_c \label{eq:geodesic-eqn2}.
\end{align}

Consider now a foliation of the spacetime in terms of null hypersurfaces $\Sigma_v$, labelled by the parameter $v$. Denote by $n_a:=-\partial_a v$ the null geodesic generators of $\Sigma_v$. Denote by $r$ an affine parameter for the geodesics generated by $n^a$, and lastly, let $x^A$ be angular coordinates such that the null geodesics are lines at constant angle: $\mathcal{L}_n x^A=0$. In these coordinates\footnote{Newman-Unti coordinates \cite{Newman:1962cia} are examples of such coordinates.} it holds true that
\begin{equation}
g^{rv}=1 \quad \text{and} \quad g^{vv}=g^{vA}=0.
\end{equation}
One may then verify that for $\tau=r$ and for a small function $f=f(v,x^A)$,  an $O(f^2)$ solution to the equations \eqref{eq:geodesic-eqn1} and \eqref{eq:geodesic-eqn2} is given by
\begin{equation}\label{eq:GE-solution}
\begin{split}
x^a &= \int^r g^{ab} p_b + z^a, \\
p_a &=  n_a -  \partial_a   f. 
\end{split}
\end{equation}
Here $z^a=z^a(v,x^A)$. 

Notice that $f=0$ yields a geodesic generated by $n^a$. The function $z^a$ determines the ingoing location of the geodesic. The linearised solution \eqref{eq:GE-solution} thus tells us that, starting from a geodesic generated by $n^a$, we may generate a family of geodesics in its neighbourhood by small deformations $f$. The geodesics \eqref{eq:GE-solution} are null and affinely parametrised by $r$.

\subsubsection*{Geodesic deviation}
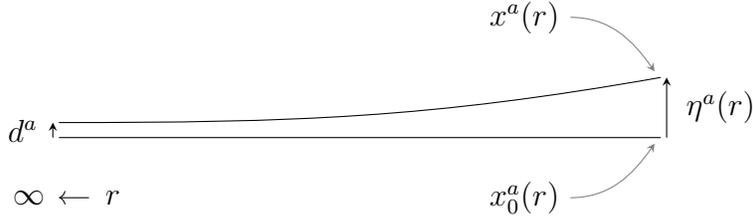
\begin{figure}[h!]
	\centering
	\begin{tikzpicture}[scale=0.8]
	%\node at (0, -0.5) {Trapped region} ;
	
	\node at (-12.6,0.12) {$d^a$};
	%\draw (0,0) circle (2);
	\draw (-12,0)  --  (-2,0);
	\draw (-12,0.25) to[out=0,in=-170](-2,1);
	\draw [<-,>=stealth] (-1.9,1)-- (-1.9,0);
	\draw [->,>=stealth] (-12.1,0)-- (-12.1,0.25);
	\draw [<-] (-12,-1) -- (-11.5,-1);
	\node at (-12.5,-1) {$\infty$};
	\node at (-11.1,-1) {$r$};
	
%	\node at (-4,2)  {$x^a(r)|_f$} ;
%	\node at  (-4,-1)   {$x^a(r)|_{f=0}$};
	\node at (-1,0.5) {$\eta^a(r)$};
	
		\draw[->,>=stealth,color=gray] ($ (-3.5,-1)$) to[out=0,in=-125] node[at start , left,color=black] {$x^a_0(r)$} ($  (-2.1,-0.1)$);
		\draw[->,>=stealth,color=gray] ($ (-3.5,2)$) to[out=0,in=125] node[at start , left,color=black] {$x^a(r)$} ($  (-2.1,1.1)$);
	
	\end{tikzpicture}
	\caption{A light ray generated by $n^a$ and a deformed light ray generated by $n^a_f$. The geodesic deviation between the two light rays at affine time $r$ is denoted by $\eta^a(r)$.
	}
	\label{fig:lambda0-1}
\end{figure}

Consider a geodesic $x^a(r)$ of the type \eqref{eq:GE-solution} and a geodesic $x^a_0(r)$ generated by the vector $n^a$. The latter is a geodesic of the type \eqref{eq:GE-solution} where $f=0$.  A \emph{deviation vector} between these geodesics is given by 
\begin{equation}\label{eq:xi-deviation}
\eta^a(r) :=x^a(r)  - x^a_0(r).
\end{equation}
This quantity (depicted in \autoref{fig:lambda0-1}) shall play a central role in the remainder of this note. Notice, however, that $\eta^a$ is not yet well-defined. Namely, the right hand side of  \eqref{eq:xi-deviation} depends on the choice of the affine parameter $r$.  We shall fix this ambiguity  in \autoref{sec:memory}.

Notice also that a generator $n_f^a$ of the null geodesic $x^a(r)$ can  be constructed from $n^a$ in the following way:
\begin{equation}\label{eq:generator-BMS-deformed}
 n_f^a=n^a + \mathcal{L}_\eta n^a,
\end{equation}
where $\mathcal{L}_\eta$ denotes the Lie derivative with respect to $\eta$.

\section{Gravitational memory}\label{sec:memory}
The geodesic deviation \eqref{eq:xi-deviation} between a light ray generated by $n^a$ and its deformation by the function $f$ can be used to detect gravitational memory at all values of the affine parameter $r$. The idea is to compare the geodesic deviation $\eta^a(r)$ at different values of $v$, which we denote by $v$ and $v'$. 

However, for a comparison of $\eta^a(r)$ at different values of $v$ to make sense in the context of quantifying the memory effect, we must impose further restrictions on the choice of the coordinates $(v,r,x^A)$ and the choices of $f$ and $z^a$. We require the following.

\begin{enumerate}
	\item[(i)] The coordinates $(v,r,x^A)$  are \emph{Newman-Unti coordinates} \cite{Newman:1962cia}. Newman-Unti coordinates are of the type $(v,r,x^A)$ above, where in addition the metric satisfies the following conditions.
	\begin{equation}\label{eq:NU-form}
	\D s^2 = {W} \D v^2 + 2  \D v \D r+ g_{AB}  (\D x^A - {V}^A \D v)(\D x^B - {V}^B \D v),
	\end{equation}
	where
	\begin{align}
	W &= -1 + O(r^{-1}),\\
	V^A &= O(r^{-2}),
	\end{align}
	and
	\begin{equation}\label{eq:asymptotic-metric}
	g_{AB} = r^2 \gamma_{AB} + {r} C_{AB} + O(1).
	\end{equation}
	Furthermore, $C_{AB}$ is traceless with respect to the round metric $\gamma_{AB}$,
	\begin{equation}\label{eq:traceless}
	\gamma^{AB}C_{AB}=0.
	\end{equation}
	In our notation, capital indices are raised with the metric $g_{AB}$, except for the indices of $\gamma_{AB}$ and $C_{AB}$, which are raised and lowered with the metric $\gamma_{AB}$.
	\item[(ii)] The functions $f$ and ingoing locations $z^a$ are independent of $v$.
\end{enumerate}

There is a physical motivation for choosing Newman-Unti coordinates. Namely, they have the property that at a large constant radius $r=r_0$, the worldlines $(v,r_0,x_0^A)$ at fixed angles $x_0^A$ are  approximately inertial observers. This means that -- together with condition (ii) -- the setup can be understood as an approximately inertial observer at past null infinity who shoots at two different times ``the same'' pair of light rays into the spacetime. The quantity $\eta^a$ can then be compared at the two different values of $v$ by a second observer in the bulk.

Assuming the requirements (i) and (ii), the quantity
\begin{equation}\label{eq:quantity}
\Delta \eta^a := \eta^a|_{v'}- \eta^a|_{v},
\end{equation}
is now well-defined at every value of $r$. We argue that this choice of $\Delta \eta^a$ quantifies the memory effect in the bulk of a spacetime.

\subsection{Memory at null infinity}\label{eq:null-infty-NU}
In order to verify that \eqref{eq:quantity} is a formulation of the memory effect, we must show that the known literature about the memory effect at null infinity is correctly reproduced in this formulation.

The memory effect was formulated by Christodoulou in \cite{Christodoulou:1991cr} in terms of a permanent relative displacement $\Delta {x}^A$ between timelike geodesics that are initially at rest. At null infinity (in $3+1$ dimensions), this displacement may be expressed as\footnote{Equation \eqref{eq:Christodoulou} is obtained by integrating twice the geodesic deviation equation between neighbouring timelike geodesics at subleading order in $r$.} \cite{Christodoulou:1991cr}
 \begin{equation}\label{eq:Christodoulou}
 \Delta {x}^A = - \frac{(\delta x_0)^B}{2r}\Delta C^A_B .
 \end{equation}
 Here $(\delta x_0)^A$ denotes the initial relative separation between the timelike geodesics and $\Delta C^A_B$ denotes the difference\footnote{An ingoing flux of radiation is equivalent to the non-vanishing of the \emph{Bondi news} defined by $N_{AB}:= \partial_v C_{AB}$.} of the \emph{asymptotic shear} at two different values of $v$. 

Our goal is now to show that our formulation yields the same result. Towards this end, consider the geodesic deviation \eqref{eq:xi-deviation} with the same choice of $z^A$ for both geodesics. Then the asymptotic expansion of the angular components of $\eta^a$ is given by\footnote{This follows from integrating
\begin{equation}
g^{AB} = \frac{1}{r^2}\gamma^{AB} - \frac{1}{r^3}C^{AB}+O(r^{-4}).
\end{equation}
}
\begin{equation}\label{eq:asympt-gd}
\eta^A =  \bigg(\frac{1}{r} \gamma^{AB}  - \frac{1}{2r^2}C^{AB}+O(r^{-3}) \bigg)\partial_B f.
\end{equation}
Next, note that the leading order deviation is 
\begin{equation}\label{eq:initial distance}
d^A := \frac{1}{r}\gamma^{AB}\partial_B f.
\end{equation}
Therefore, we may express the change \eqref{eq:quantity} in the geodesic deviation \eqref{eq:asympt-gd} as
\begin{equation}\label{eq:memory-infinity}
\Delta \eta^A= - \frac{d^B}{2r} \Delta C^A_B   +O(r^{-3}).
\end{equation}
Equation \eqref{eq:memory-infinity} is the same as \eqref{eq:Christodoulou}. This shows that our proposal captures the memory effect at null infinity.

\subsection{Memory in the bulk}
Here, we quantify the memory of an accreting black hole. The purpose of this is to illustrate that our method is applicable in the bulk of a spacetime.

\begin{figure}[h!]
	\centering
	\begin{tikzpicture}[ scale=0.6]
	%	\path % Four corners - invisible path
	%	(-1.5,-1.5)  coordinate-- (-5,5)  --  (-1,5)  coordinate[label=90:$i^+$] -- (2,2)  coordinate[label=0:$i^0$] -- cycle
	;
	\coordinate (i-) at (0,0) {};
	\coordinate (i0) at (8,8) {};
	\node (shift) at (1,1) {};
	\node (rd-shift) at (-1,1) {};
	%	\node (I)    at ( -1,2.5)   {$\mathcal{M}(\Delta)$};
	\draw[] (i-)   --  (i0);
	%\draw ($ (i-) + 0.5*(shift) $) -- ($ (i0) - 0.5*(shift) $);
		\draw[->,>=stealth] ($(i0) + (1,0.5)$) to[out=-90,in=0] node[at start,above ]{$\mathcal{I}^-$} ($(i0) + (0,-0.5)$);
	
	%\draw ($ (i-) + 0.5*(shift) $) -- ($ (i0) - 0.5*(shift) $);
	%	\draw ($(i-)+2*(-1,1)+1*(-1,1)$)  to[out=5,in=-160] node [at end,above left]    {${}^{(3)}\!B$} ($(i-)+3*(1,0)+7*(0,1)+1*(-1,1)$);
	\draw[dashed,color=gray] ($(i-)+2*(shift)$) -- ($(i-)+2*(shift)+4*(-1,1)$);
	\draw[dashed,color=gray] ($(i-)+1.8*(shift)$) --  ($(i-)+1.8*(shift)+4*(-1,1)$);
	
	\draw[dashed,color=gray] ($(i-)+4*(shift)$) -- ($(i-)+4*(shift)+4*(-1,1)$);
	\draw[dashed,color=gray] ($(i-)+4.2*(shift)$) -- ($(i-)+4.2*(shift)+4*(-1,1)$);
	
	\node[ left] (I)    at ($(i-)+1.9*(shift)+4*(-1,1)$)  {$\eta^A|_{v}$};
	\node[above right] (II)    at ($(i-)+4.1*(shift)+4*(-1,1)$)  {$\eta^A|_{v'}$};
	
	\draw[->, thick,>=stealth] ($(i-)+1.8*(shift)+4*(-1,1)$) -- ($(i-)+2*(shift)+4*(-1,1)$);
	\draw[->, thick,>=stealth] ($(i-)+4*(shift)+4*(-1,1)$) -- ($(i-)+4.2*(shift)+4*(-1,1)$);
	
	\draw[->, thick,>=stealth] ($(i-)+1.8*(shift)-0.1*(rd-shift)$) -- ($(i-)+2*(shift)-0.1*(rd-shift)$);
	\draw[->, thick,>=stealth] ($(i-)+4*(shift)-0.1*(rd-shift)$) -- ($(i-)+4.2*(shift)-0.1*(rd-shift)$);
	
	\node[below right ] (Ia)    at ($(i-)+1.9*(shift)$)  {$d^A$};
	\node[ below right] (IIa)    at ($(i-)+4.1*(shift)$)  {$d^A$};
	%	\draw[black,fill=black] (-0.02,6.02) circle (.5ex) node [right] {$B_v$};
	%	\draw (-0.02,6.02) node [right] {$B_v$};
	
	\draw[->,decorate, decoration={snake}] ($(i-)+3*(shift)$)  -- node[at start, right, rotate=-45] {$ v=v_0$} ($(i-)+3*(shift)+9.1*(-1,1)$);
		
	%%%----Black Hole---%%%
	\draw[ ] ($(i-)+8*(rd-shift)+(1.5,-0.5)$)  to[out=45,in=-135]  node [at start, left, rotate=0] {$r=2m$}  node [midway,  above left] {$\mathcal{H}$} ($(i-)+8*(rd-shift)+(2,0)+(2,2)$) -- ($(i-)+8*(rd-shift)+(2,0)+(2,2)+(0.5,-0.5)$)-- node [at end, right, rotate=0] {$r=2m+2\mu+\frac12 \overset{\circ}{D}{}^2 C$}  ($(i-)+8*(rd-shift)+(2,0)+(4.5,4.5)+(0.5,-0.5)$) ; 
	
	\end{tikzpicture}
	\caption{A black hole subject to a shell of null matter entering the spacetime at $v=v_0$. The geodesic deviation between a pair of light rays before $v_0$ and ``the same'' pair of light rays after $v_0$ may be compared as a way of detecting gravitational memory in the exterior of the black hole. Here ``the same'' is given a meaning by the conditions (i) and (ii) in the beginning of this section.}
	\label{fig:shockwave}
\end{figure}
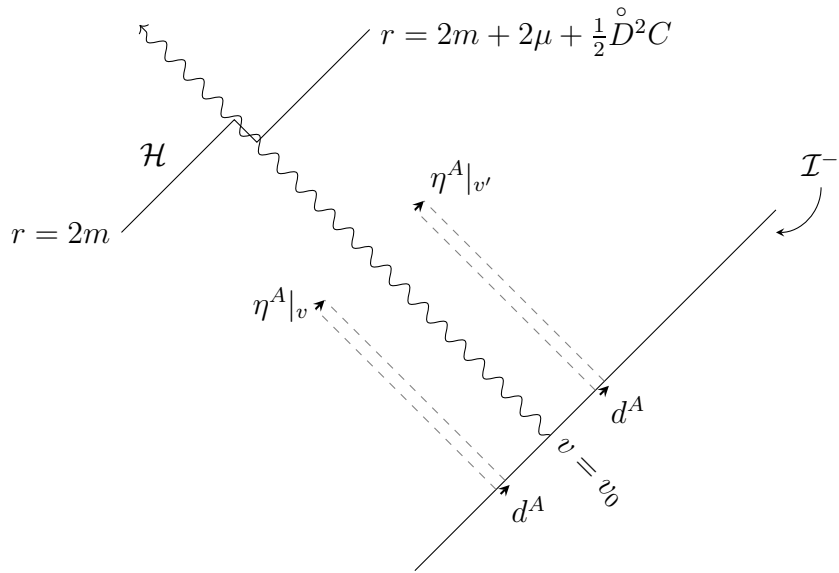

We consider a Schwarzschild black hole subject to an ingoing shell of null matter, given by the metric $g  = g_0 + h$. Here $g_0$ denotes the Schwarzschild metric 
\begin{equation}
\D s^2_0 = -\bigg(1-\frac{2m}{r}\bigg)\D v^2 + 2\D v \D r + r^2 \gamma_{AB} \D x^A \D x^B,
\end{equation}
and $h$ is a perturbation that describes the shell of matter, given by  \cite{Hawking:2016sgy}
\begin{align}
%h_{vv} &= \theta(v-v_0) \bigg(\frac{2\mu}{r}- \frac{m \overset{\circ}{D}{}^2 C}{r^2}\bigg),\\
%h_{vA} &= \theta(v-v_0)\partial_A\bigg[ \bigg(1-\frac{2m}{r}+\frac12 \overset{\circ}{D}{}^2\bigg)C\bigg],\\
h_{AB} &= \theta(v-v_0) r C_{AB},
\end{align}
where 
\begin{equation}
C_{AB} = -2\bigg(\overset{\circ}{D}_A \overset{\circ}{D}_B C - \frac12  \gamma_{AB} \overset{\circ}{D}{}^2 C\bigg).
\end{equation}
 Here $\overset{\circ}{D}$ denotes the covariant derivative with respect to $\gamma_{AB}$ and $\theta$ denotes the Heaviside step function. The remaining components\footnote{The remaining components of the perturbed metric $h_{ab}$ are given by\begin{align}
 	\label{eq:hvv} h_{vv} &= \theta(v-v_0) \bigg(\frac{2\mu}{r}- \frac{m \overset{\circ}{D}{}^2 C}{r^2}\bigg),\\
 	\label{eq:hvA} h_{vA} &= \theta(v-v_0)\partial_A\bigg[ \bigg(1-\frac{2m}{r}+\frac12 \overset{\circ}{D}{}^2\bigg)C\bigg].
 	\end{align}} of $h$ are given by $h_{vr}=h_{rA}=0$, \eqref{eq:hvv} and \eqref{eq:hvA}. The linearised solution $h$ is constructed so that, before and after $v_0$, the metric is diffeomorphic to the Schwarzschild geometry with masses $m$ and $m+\mu$ respectively. The setup is depicted in \autoref{fig:shockwave}.

One may compute that for $v>v_0$,
\begin{equation}\label{eq:geodesic-deviation-v0}
\eta^A = \bigg(\frac{1}{r} \gamma^{AB} -\frac{1}{2r^2}C^{AB} \bigg)\partial_B f.
\end{equation}
And for $v<v_0$, $\eta^A$ is given by \eqref{eq:geodesic-deviation-v0} with $C_{AB}=0$. Consider then the change
\begin{equation}\label{eq:difference-Schwarzschild}
\Delta \eta^A  = - \frac{1}{2r^2}\Delta C^{AB}\partial_Bf.
\end{equation}
The deviation  $\Delta \eta^A$ quantifies the displacement memory effect in the bulk\footnote{A similar statement was made in \cite{Hawking:2016sgy}, although without the interpretation provided in the present paper.}. Namely, it\footnote{The apparent discrepancy between the asymptotic behaviour of \eqref{eq:memory-infinity} and \eqref{eq:difference-Schwarzschild} is due to  \eqref{eq:initial distance}.} is defined for all values of $r$ and it vanishes\footnote{This would not be true without the conditions (i) and (ii) in the beginning of this section.} when $v$ and $v'$ are taken both before or both after the incoming shell.

Note that in general, in addition to $\Delta \eta^A$, one may also consider the deviation $\Delta \eta^r$ in the longitudinal direction.

\section{BMS symmetries}\label{sec:BMS}
Generators of \emph{BMS symmetries}\footnote{The original references are \cite{Bondi:1962px,Sachs:1962wk,Sachs:1962zza}. See \cite{Ashtekar:2014zsa,Barnich:2011mi,Strominger:2017zoo} for reviews.} are vector fields that preserve the asymptotically flat boundary conditions at null infinity. A connection between the gravitational memory effect and BMS generators at null infinity was discovered in \cite{Strominger:2014pwa}. The relation is that a change in the deviation between geodesics may be understood as the action of a BMS supertranslation. Here we show that the  formulation of the gravitational memory effect proposed in \autoref{sec:memory} generalises this connection to the bulk of a spacetime. 

Usually, BMS symmetries are only considered in the asymptotic region. Their action in the bulk is considered arbitrary and therefore physically irrelevant. However, when a gauge such as Bondi or Newman-Unti gauge has been fixed, BMS generators have a unique extension into the bulk. 

We consider  BMS generators in Newman-Unti gauge \cite{Newman:1962cia,Barnich:2011ty} and observe that they are given by the geodesic deviation vector  \eqref{eq:xi-deviation}, where $x^a_0(r)$ is a geodesic of the type \eqref{eq:GE-solution} where $f=0$ and $z^a=z_0^a$, and $x^a(r)$ is a geodesic of the type \eqref{eq:GE-solution} where
\begin{equation}\label{eq:BMS-anc}
\begin{split}
f &= T(x^A) + \frac12 v \overset{\circ}{D}_A Y^A,\\
z^v &= f +z_0^v,\\
z^r &= \frac12 \overset{\circ}{\Delta}f+z_0^r, \\
z^A &= Y^A+z_0^A.
\end{split}
\end{equation}
 Here $T$ is a function depending only on $x^A$, referred to as a \emph{supertranslation}, and $Y^A$ is a conformal Killing vector of $\gamma_{AB}$, referred to as a \emph{superrotation}. The operator  $\overset{\circ}{\Delta}:=\overset{\circ}{D}_A\overset{\circ}{D}{}^A$ is the spherical Laplacian. This observation tells\footnote{It could be interesting to see if this observation has implications for defining BMS symmetries in higher dimensions \cite{Hollands:2016oma}.} us that, in Newman-Unti gauge, BMS vector fields are themselves the geodesic deviation between a light ray generated by $n^a$ and a light ray generated by a BMS deformation of $n^a$. Upon closer inspection of \eqref{eq:BMS-anc}, the bulk memory in \eqref{eq:quantity} may be understood as the action of a supertranslation. This extends the connection between supertranslations and gravitational memory \cite{Strominger:2014pwa} to the bulk.

BMS generators in Bondi gauge \cite{Barnich:2011mi}, where the affine radial coordinate is replaced by the \emph{luminosity} or \emph{areal distance}, are not related to the gravitational memory effect in this way\footnote{Consider the metric \eqref{eq:NU-form} given by $W = -1$, $g_{AB} = r^2\gamma_{AB}$ and $V^A = V^A(r)$. Then the Newman-Unti coordinate $r$ is also the Bondi coordinate $r$. The BMS generators in the Bondi and Newman-Unti gauges are related by $\xi^a_B = \xi^a_{NU} + F n^a$, where $F$ is a function of the coordinates including $r$. Since geodesic deviation vectors are related by a function $F$ which is independent of the affine parameter $r$, this shows that BMS generators in Bondi gauge are in general not a geodesic deviation.}.

\section*{Acknowledgements}
I thank Lasha Berezhiani, Dieter L\"ust, Daniel Kl\"awer, David Osten, Piotr Witkowski and in particular Sebastian Zell for useful discussions and feedback on the draft.

\appendix

\end{spacing}

\bibliographystyle{JHEP}

\bibliography{references}

\end{document}